\documentclass[10pt, a4paper, twocolumn]{article}

\usepackage{authblk}
\usepackage{geometry}
\usepackage{mathpazo}
\usepackage{cite}
\usepackage{amsmath,amssymb,amsfonts}
\usepackage{algorithmic}
\usepackage{graphicx}
\usepackage{textcomp}
\usepackage{xcolor}
\usepackage{float}
\usepackage{nicefrac}
\usepackage{xspace}
\newcommand{\quic}{\textsc{Quic}\xspace}
\usepackage{pgffor}
\usepackage[hyphens]{url}
\usepackage{siunitx}
\DeclareSIUnit\TPS{TPS}

\newcommand{\code}[1]{{\fontfamily{cmss}\selectfont\textcolor{gray}{#1}}}

\usepackage[hidelinks]{hyperref}
\usepackage[capitalize]{cleveref}

\def\BibTeX{{\rm B\kern-.05em{\sc i\kern-.025em b}\kern-.08em
    T\kern-.1667em\lower.7ex\hbox{E}\kern-.125emX}}

\newcommand{\ms}[1]{%
    \relax\ifmmode
        \mathord{\mathcode`\-="702D\it #1\mathcode`\-="2200}%
    \else
        {\it #1}%
    \fi
}

\begin{document}

\title{On the Bandwidth Consumption of Blockchains%
}

\author[1]{Andrei Lebedev}
\author[1,2]{Vincent Gramoli}
\affil[1]{The University of Sydney}
\affil[2]{Redbelly Network}
\date{}

\maketitle

\begin{abstract}
With the advent of blockchain technology, the number of proposals has boomed. The network traffic imposed by these blockchain proposals increases the cost of hosting nodes. Unfortunately, as of today, we are not aware of any comparative study of the bandwidth consumption of blockchains.

In this paper, we propose the first empirical comparison
of blockchain bandwidth consumption. To this end, we measure the network traffic of blockchain network nodes of five blockchain protocols: Algorand, Aptos, Avalanche, Redbelly and Solana. We study the variation over time, differentiate the receiving and sending traffic and analyze how this traffic varies with the number of nodes and validators. 

We conclude that the transport protocol is the main factor impacting the network traffic, segregating node roles helps reduce traffic and different blockchains are differently impacted by the network size.
\end{abstract}

\section{Introduction}

With the advent of blockchain technology, the number of proposals has boomed over the past decade.  There exist various layer-1 blockchains offering different guarantees and performing differently at large scale. 
Recent studies~\cite{GGLNV23,GGLV25} have shown that the performance and robustness of some blockchains is quite far from the promises claimed by their designers.
The problem stems mainly from a misunderstanding of these blockchains' underlying networking protocols. %
The impact of these misunderstandings is so important that
Avalanche~\cite{rocket_scalable_2020} and Solana~\cite{Yak21} were even shown to stop globally when some network messages get delayed~\cite{GGLV25}.

The misunderstanding of these networking protocols also presents economic drawbacks.
Modern high-throughput architectures have shifted validator economics from static capital expenditures to variable operational burdens, where egress consumption acts as a primary cost driver. For instance, the prohibitively high egress tariffs of standard cloud providers can render high-performance nodes economically non-viable, potentially incurring monthly costs exceeding US\$\num{8000} for a single validator due to misaligned billing models~\cite{Hiv26}. This structural inefficiency necessitates a strategic pivot toward bare-metal providers with unmetered allowances to avoid the ``hyperscaler egress trap''~\cite{Wev25}.

As of today, there are no studies
comparing
the network traffic of blockchains.
Even though it is well known that some blockchains (e.g.,  Solana~\cite{Yak21}) favor redundancy (e.g., through erasure coding) to cope with packet losses while others (e.g., Redbelly~\cite{CNG21}) adopt different communication patterns between validator and non validator nodes,  the bandwidth consumption of these blockchains remain unclear. Understanding the network traffic is, however, crucial to %
improve
performance and robustness of layer-1 blockchains by reducing the network congestion to make them scale or to replicate the data that are key to their robustness. %

In this paper, we propose the first empirical comparison of blockchain bandwidth consumption.
To this end, we build upon the series of work around the Diablo performance benchmark~\cite{GGLNV23} and the {\sc Stabl} fault tolerance benchmark~\cite{GGLV25} to measure the network traffic of blockchain network nodes in various situations: (i)~while receiving and sending messages; (ii)~before,  during and after the network receives transactions and (iii)~as the network size grows both in terms of nodes but also validator nodes.

Using this black-box approach, we deploy five different blockchain protocols, namely Algorand~\cite{GHM17},  Aptos~\cite{noauthor_aptos_2022}, Avalanche~\cite{rocket_scalable_2020}, Redbelly~\cite{CNG21}, and Solana~\cite{Yak21}.
We make the following observations:

\begin{enumerate}
\item The dominant factors of network traffic are the transport protocol (polling vs WebSockets) and the block propagation strategy (full block download vs.  hash comparison) more than the transaction size.
\item Segregating roles between different types of nodes helps reduce the 
network traffic by reducing network traffic between nodes of different types. 
\item Solana network traffic depends on the network size,  Algorand and Redbelly network traffic increases with the validator sets and Aptos and Avalanche network traffic increases with both the number of nodes and validators.
\end{enumerate}

The paper is organized as follows.  In \cref{sec:bg}, we present the background. 
 In \cref{sec:exp},  we present the experimental settings and our methodology. 
 In \cref{sec:bandwidth},  we present the variety of bandwidth consumption of the five blockchain protocols.
 In \cref{sec:routes},  we study the distribution of network traffic over different pairs of nodes.
In \cref{sec:time}, we compare the bandwidth consumption before, during and after reception of transactions.
In \cref{sec:scale}, we study the impact of the number of nodes and validators on the bandwidth consumption. 
In \cref{sec:tps}, we study the impact of the sending rate on the bandwidth consumption.
We present the related work in \cref{sec:rw} and  we conclude in \cref{sec:conclusion}.

\section{Background and Blockchain Networks}\label{sec:bg}

In this section, we list the characteristics of each tested blockchain network protocol.

\subsection{The Algorand network}\label{sec:algorand}

Algorand~\cite{GHM17} is a blockchain protocol that shuffles participants via cryptographic sortition to enhance security. More precisely, it uses Verifiable Random Functions (VRFs) to randomly select participants for specific roles in the consensus execution. 
In Algorand, nodes communicate with a gossip-based protocol where each node validates each message before relaying it and sends it at most once to each other node~\cite{CGT19}. To this end, each node maintains one TCP connection per node in its neighborhood, which offers WebSockets over HTTP.

\subsection{The Avalanche network}\label{ssec:avalanche}

Avalanche, based on the Snowflake consensus protocol~\cite{rocket_scalable_2020}, is a probabilistic blockchain that requires, by default, a proportion of the nodes that collectively own at least 80\% of the total stake to be online.
In Avalanche, nodes communicate over TCP  and exploit throttling to limit their resource usage.
More specifically, messages and connections are rate-limited~\cite{noauthor_avalanchego_nodate} to cap the amount of CPU, disk, bandwidth, and message handling that other nodes consume.
Avalanche uses a dynamic proposer selection algorithm to manage network load and block production. After each parent block, it pseudo-randomly selects an ordered list of potential proposers for the next block height, weighted by stake and using a seed derived from the parent block's height and chain ID. Each proposer is assigned a minimum delay based on their position in the list before they can propose a block. If no proposer acts within the cumulative delay period, any active validator may propose. Blocks within the assigned windows must include valid signatures from the designated proposer.

\subsection{The Aptos network}\label{sec:aptos}

Aptos~\cite{noauthor_aptos_2022} is a leader-based blockchain that uses TCP and builds upon a variant of the \emph{Practical Byzantine Fault Tolerant (PBFT)} consensus protocol~\cite{castro_practical_1999}
featuring 
a view-change mechanism with a quadratic communication complexity 
and inherits its cubic communication complexity reached when the leader is faulty or the network is unstable. 

It is well-known that leaders act as bottlenecks in leader-based consensus protocols~\cite{VG20}, like AptosBFT, and that classic blockchains suffer from redundant dissemination of the same transactions first outside and then within blocks~\cite{THG23,TG24}. 
To cope with these limitations, %
Aptos features the Quorum Store optimization of Narwhal~\cite{danezis_narwhal_2022} to decouple metadata ordering from payload dissemination. 
In addition, Quorum Store is designed for parallel execution by all validators. As outlined in the documentation~\cite{aip-26}, validators repeat the following steps in parallel: \textit{(1)}~Pull transactions from the mempool; \textit{(2)}~Arrange transactions into batches based on gas price and select an expiration time for each batch; \textit{(3)}~Broadcast batches to all other validators; \textit{(4)}~Persist received batches, sign their digests, and send back signatures; and \textit{(5)}~Collect signatures from more than $\nicefrac{2n}{3}$ nodes to form a \emph{proof-of-store}.
It allows validators to asynchronously broadcast transactions, offloading the leader's network interface during the consensus protocol execution~\cite{apt-quorum-store}.

\subsection{The Redbelly network}

Redbelly Blockchain~\cite{CNG21} is a scalable blockchain built on the Democratic Byzantine Fault Tolerant (DBFT) consensus algorithm~\cite{CGLR18} that is \emph{leaderless} (i.e., non leader-based) and deterministic, and works in a partially synchronous environment.
To enhance scalability further, Redbelly uses a collaborative approach, %
appending a superblock with as many valid proposed blocks as possible. This way the number of transactions per appended block can grow linearly with the number of nodes~\cite{CNG21}.

Redbelly's nodes communicate using TCP
and features a Scalable variant of the EVM, called SEVM.
It was shown to perform well under realistic dApps particularly in a large geo-distributed environment when compared to other modern blockchains~\cite{THG23}.

\subsection{The Solana network}\label{ssec:solana}

Solana~\cite{Yak21} is a leader-based blockchain that may fork. In order to determine whether a transaction is committed, Solana requires 30 additional blocks to be appended after the transaction's block.
Nodes communicate over the \quic network protocol~\cite{IETF21} to exchange transactions.
Nodes split blocks into chunks that they disseminate in a hierarchical structure, called Turbine~\cite{anza-turbine}, through UDP. %

\section{Experimental Settings}\label{sec:exp}

In this section, we explain how we deploy blockchain protocols and measure their bandwidth consumption.

\subsection{Distributed system setup}

Our experimental setup consists of 
a distributed system of 25 VMs running Ubuntu 24.04.1 LTS on top of a Proxmox cluster of physical servers, each equipped with 4x AMD Opteron 6378 16-core CPUs running at \qty{2.40}{\giga\hertz}, \qty{256}{\giga\byte} of RAM, and 10 GbE NICs. 
Each experiment runs a blockchain protocol with 5 client VMs and 20 blockchain node VMs.
For the blockchain protocols, we used 
Algorand v3.27.0, Aptos v1.25.1, Avalanche C-Chain v1.12.1, Redbelly v0.36.2 and Solana Agave v2.0.20.

This 25-node setup is justified by recent studies~\cite{LG23} that confirmed
that evaluations at small scale in controlled environments can accurately reproduce performance trends observed in geo-distributed settings.  We define $N$ as the total number of blockchain nodes in the network, and $V$ as the size of the subset of validators participating in consensus such that $N, V \in \{5, 10, 15, 20\}$, $V \le N$.

All experiments follow a fixed timeline: a \qty{100}{\second} ``Setup'' phase with no transactions, a \qty{100}{\second} ``Workload'' phase where transactions are sent, and a \qty{100}{\second} ``Cooldown'' phase with no transactions, allowing remaining transactions to commit. During the Workload phase, the 5 clients send transactions to the first 5 validators. The target load is distributed equally among clients (e.g., \qty{40}{\TPS} each for a total of \qty{200}{\TPS}). Each transaction is sent to a single node, which is then queried for finality using block streaming or polling.

The resources of each VM mimics intentionally the resources of a 
commodity computer run by an individual in a blockchain network.
Note that this specification is lower than what some blockchains typically recommend, including 
Aptos~\cite{apt-reqs}, 
Avalanche~\cite{ava-fundamentals} or 
Redbelly~\cite{redbelly-reqs}, however, strict hardware requirements on remote nodes remain hard to enforce and a unique configuration is necessary for our comparison.

\begin{figure*}
    \centering
    \includegraphics{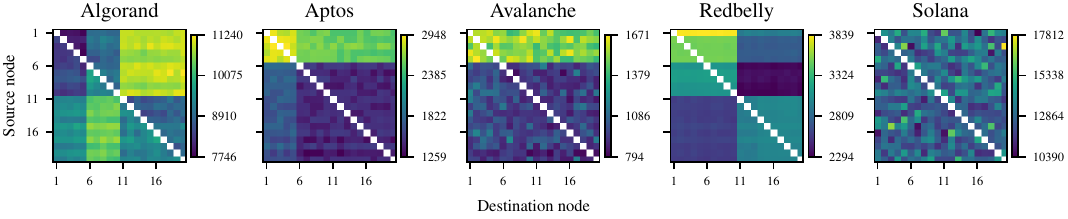}
   \caption{Heatmaps $M_{i,j}$ of the bandwidth used in \unit{\kibi\byte} between sending node $n_i$ and receiving node $n_j$ for each blockchain, 20 nodes, 20 validators.\label{fig:total-20-20}}
\end{figure*}

\subsection{Measuring peer-to-peer bandwidth}

We implemented a fine-grained bandwidth monitoring system to capture traffic usage between blockchain nodes and clients. Our approach provides pairwise measurements of the traffic exchanged between each node in the network. The system utilizes {\sc Stabl} observer processes running on blockchain VMs and relies on the Linux \code{iptables} firewall infrastructure to perform non-intrusive packet accounting.

For each node under observation, we programmatically install a set of \code{iptables} rules. These rules create custom accounting chains that contain a specific rule for every other peer in the experiment. Each rule is configured to match packets based on their source (for incoming traffic) or destination (for outgoing traffic) IP address.

A monitoring script then periodically queries the byte counters associated with each of these per-peer rules and immediately resets them to zero. This process yields a time series where each data point represents the average transmission (TX) and reception (RX) rate over the preceding interval, allowing us to precisely analyze network behavior.

\section{Varying Consumption}\label{sec:bandwidth}

In order to illustrate how blockchains consume bandwidth, we present heatmaps with colors representing bandwidth usage. We then compare the transaction size produced by each blockchain.

\subsection{Bandwidth consumptions as heatmaps}

Figure~\ref{fig:total-20-20} depicts one heatmap per blockchain protocol for nodes that all act as validators where the color of cell $M_{i,j}$ represents the bandwidth consumed by node $n_i$ when sending to node $n_j$. 
A warmer color (e.g., yellow) thus represents a higher bandwidth usage than a colder color (e.g., dark blue) while the white color indicates zero or negligible traffic.
Note that nodes $n_1, \dots, n_5$ are the nodes receiving transaction requests from the clients, while nodes $n_6, \dots, n_{20}$ may receive messages from other nodes but not transactions directly from clients. 
With a maximum of \qty{17812}{\kibi\byte}, Solana consumes more bandwidth than the other tested blockchains, namely Algorand, Aptos, Avalanche and Redbelly.
In particular, %
Algorand, the second most bandwidth consuming blockchain uses a maximum of \qty{11240}{\kibi\byte}. This 58\% increase compared to Algorand can be due to several factors: more metadata sent per transaction or higher duplication of the same information.

The reason is probably that Solana exploits
erasure coding in order to maximize dissemination of information despite failures~\cite{helius-turbine}. Its lack of fault tolerance, already previously observed~\cite{GGLV25}, probably motivated this design decision.
Erasure coding relies on sending additional data in order to reconstruct the relevant information in case of partial loss. %
The other blockchains do not consume as much traffic likely because they do not use erasure coding.

\subsection{Transaction sizes}

In order to exclude other factors that could have invalidated our hypothesis that Solana uses more bandwidth due to erasure coding, we measured empirically the size of transactions sent by each blockchain. After all, it was  previously noted that distributed ledger technologies like Corda send transactions as large as \qty{8}{\kibi\byte}, which is an overkill~\cite{HSVBG24}.

\begin{figure}[H]
  \includegraphics{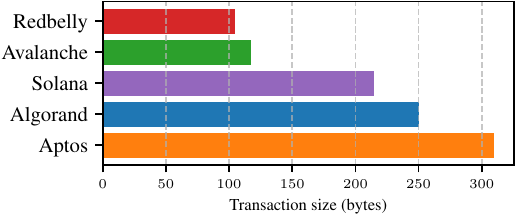}
  \caption{Transaction size per blockchain.\label{fig:transaction-sizes}}
\end{figure}

\cref{fig:transaction-sizes} compares the transaction sizes of the different blockchains. It shows in particular that, even though Solana uses more bandwidth as discussed above in \cref{sec:bandwidth}, it does not use the largest transactions. Actually, Aptos' transaction size is \qty{310}{\byte} compared to Solana's transaction size of \qty{215}{\byte}, Aptos' transactions are 44\% larger than Solana's. Finally, note that even Algorand's transactions, with a size of \qty{250}{\byte}, are larger than Solana's transactions as well. As a result, the bandwidth usage of Solana is not a consequence of generating excessively large transactions.

\section{Consumption Skewness over Routes}\label{sec:routes}

A high bandwidth usage does not necessarily induce a detrimental performance, especially when the bandwidth consumption is well-balanced over multiple routes. 
In fact, previous works have shown that balancing 
an amount of information that is quadratic in the number of network nodes over a quadratic number of routes of this network could be more efficient than reducing this amount of information to a linear factor but sent across the same routes~\cite{VG23}.
It is thus important to understand how a blockchain protocol balances its bandwidth consumption over a network.

\subsection{Some nodes exchange more data than others}

In the Aptos heatmap of \cref{fig:total-20-20}, 
we can see that some nodes of the Aptos network seem to consume more bandwidth than other nodes of the same network, and traffic is heavily unbalanced across nodes.
Five nodes, the ones that receive the transactions sent by the clients, send more messages than all the other nodes as indicated by the top five rows in light colors. They also send more messages to themselves than to the rest of the nodes, as indicated by the yellow 5-by-5 sub-matrix of the top left corner of the heatmap. 
This is due to its Quorum Store optimization~\cite{aip-26} that puts more load on the receivers of transactions than on the rest of the network by requiring them to collect signatures. More precisely, these validator nodes have to sign transaction batch digests and 
collect the produced signatures from a quorum of blockchain nodes to form a ``proof-of-store''. This signature collection puts inevitably more bandwidth pressure on the nodes responsible for signing.

In the Solana heatmap of \cref{fig:total-20-20}, the color shows that there is no clear per-node distinction as no column or row stands out. As a result, the bandwidth consumption of Solana appears generally more balanced than the one of Aptos. 
It is interesting to note that Solana, which consumes a lot of bandwidth as we showed in \cref{sec:bandwidth}, manages to balance the load pretty well.

Finally, the Algorand heatmap of \cref{fig:total-20-20} shows some imbalance in that the second half of the nodes $n_{11}, \dots, n_{20}$ receive and send more messages than others.

\subsection{Some nodes send more than they receive in Avalanche and Redbelly} 

Another interesting dimension to consider is the level of unbalance between sending and receiving traffic.  The nodes receiving transactions could either consume more bandwidth by propagating the information or, instead, 
they could consume less than the nodes agreeing on which block to append. For example, 
the Avalanche heatmap of \cref{fig:total-20-20} shows
that the nodes of Avalanche that receive transactions generate more traffic than what they receive.

The Avalanche heatmap of \cref{fig:total-20-20}
shows the five top rows in lighter color than the five left columns. This indicates that the five Avalanche nodes that receive transactions send more data than they receive. 
By contrast with Aptos, they do not need to send more data to themselves than to the rest of the network.
This is explained
by the fact that they have to propagate the transactions they receive to the rest of the network without needing to collect signatures from each other.

\begin{figure}
\includegraphics{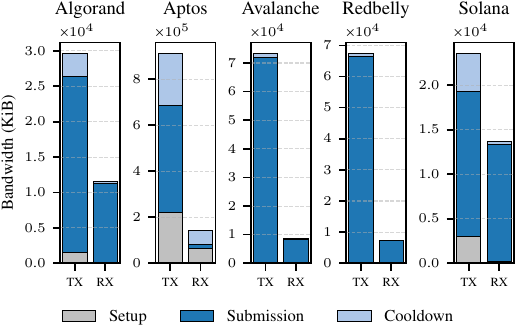}
  \caption{Client-side network traffic (TX and RX) stacked by experimental phase. Y-axis scales differ across subplots to accommodate variations in magnitude.\label{fig:client-traffic-breakdown}}
\end{figure}

The Redbelly heatmap of \cref{fig:total-20-20} shows one particular node sending more messages than others.
This is explained by having a weak coordinator that sends a particular message in the first round of the DBFT binary consensus protocol~\cite{CGLR18}. Note that this coordinator is weaker than a leader in that the consensus terminates even when it is faulty.
Generally, the first five nodes send more messages than others because they forward the transactions that this experiment sends them.
Finally, the remaining differences could be due to the node placement and the reordering of messages, requiring some nodes to request the batch of proposed transactions from other nodes.

\subsection{The dissemination reliability of Solana requires more traffic}

Bandwidth consumption can have some advantages, for example, when redundant information copes with packet losses.
The trade-off between bandwidth efficiency and dissemination reliability is particularly apparent in Solana. Previous research has highlighted Solana's ability to tolerate packet losses better than other blockchains~\cite{GGLV25}. Our experiments quantify the cost of this dissemination reliability.

The Solana heatmap of \cref{fig:total-20-20} shows the bandwidth consumption is well balanced and very high.
Unlike other protocols that attempt to minimize redundant transmissions (resulting in the dark blue or empty regions seen in Algorand or Redbelly), Solana appears to utilize a ``flood'' or highly redundant propagation mechanism, likely related to its Turbine block propagation protocol and erasure coding schemes. While this results in significantly higher total bandwidth consumption, it ensures that data is recoverable and available to all nodes, even in the presence of network failures.

\section{Consumption Skewness over Time}\label{sec:time}

In order to better understand the bandwidth usage, we measured the bandwidth consumed at the three different stages of our experiments (Setup, Submission, and Cooldown) as defined in \cref{sec:exp}.

\cref{fig:client-traffic-breakdown} differentiates the Transmitted (TX) data, which are sent from nodes to clients, from the Received (RX) data, which are sent from clients to nodes, when 5 clients submit \num{19995} transactions (\num{3999} each) to 5 nodes.

\subsection{Impact of communication protocols on idle traffic}
A distinct disparity is visible in the overhead traffic during the non-submission phases (Setup and Cooldown). Aptos exhibits significantly higher bandwidth consumption during these idle periods compared to other blockchains. For instance, during Phase 0 (Setup), Aptos nodes transmitted \qty{218856}{\kibi\byte} of data, whereas Avalanche and Redbelly transmitted only \qty{298}{\kibi\byte} and \qty{11}{\kibi\byte}, respectively.

This massive overhead in Aptos is attributed to its client communication architecture. While Avalanche, Redbelly, and Solana utilize WebSocket streaming to push updates to clients efficiently, Aptos clients rely on polling. Consequently, Aptos clients repeatedly request data even when blocks are empty, resulting in substantial bandwidth usage (\qty{\approx 227000}{\kibi\byte} in Phase 2) despite the absence of new transaction submissions.

\subsection{Block verification and data efficiency}
During the Submission phase (Phase 1), a clear divergence in data efficiency emerges between Solana and the chains compatible with the Ethereum  Web3 WebSocket API (Avalanche, Redbelly) that use the same methods to send and listen to blocks as Ethereum.

\paragraph{Full block transmission}
These protocols exhibit a high ratio of TX to RX traffic. For example, Redbelly received \qty{7314}{\kibi\byte} of transaction data (RX) but transmitted \qty{66511}{\kibi\byte} (TX) back to the clients. This amplification occurs because these protocols require the node to broadcast the entire block (containing transaction bodies and metadata) to every client for verification. Even though a client only needs transaction hashes to confirm their finality, it must download the full block payload.

\paragraph{Signature-based verification}
Solana demonstrates a more balanced traffic profile during submission (RX: \qty{13099}{\kibi\byte} vs. TX: \qty{16325}{\kibi\byte}). Despite Solana having a relatively large raw transaction size (\qty{215}{\byte}) compared to Redbelly (\qty{105}{\byte}) or Avalanche (\qty{117}{\byte}), its outgoing traffic is significantly lower. This efficiency stems from Solana's verification mechanism; clients requesting block confirmation do not need to download the full block body. Instead, they request only the hashes or signatures of committed transactions and compare them against their locally stored payloads. This selective data retrieval significantly reduces the egress bandwidth required by the nodes.

In summary, while the raw transaction sizes (ranging from \qty{105}{\byte} on Redbelly to \qty{310}{\byte} on Aptos, as displayed in \cref{fig:transaction-sizes}) play a role in bandwidth usage, the dominant factors influencing network traffic are the choice of transport protocol (polling vs. WebSockets) and the block verification strategy (full block download vs. hash comparison).

\section{Consumption Skewness over Scale}\label{sec:scale}

In this section, we analyze the scalability of the bandwidth consumption, denoted as $S$, as a function of the number $N$ of nodes and the number $V$ of validators across the different blockchains. Interestingly, we identify that the bandwidth of Aptos, Avalanche and Solana increases with $N$ and the bandwidth of Algorand, Aptos, Avalanche and Redbelly increases with $V$.

\begin{figure*}[!htb]
    \centering
    \includegraphics{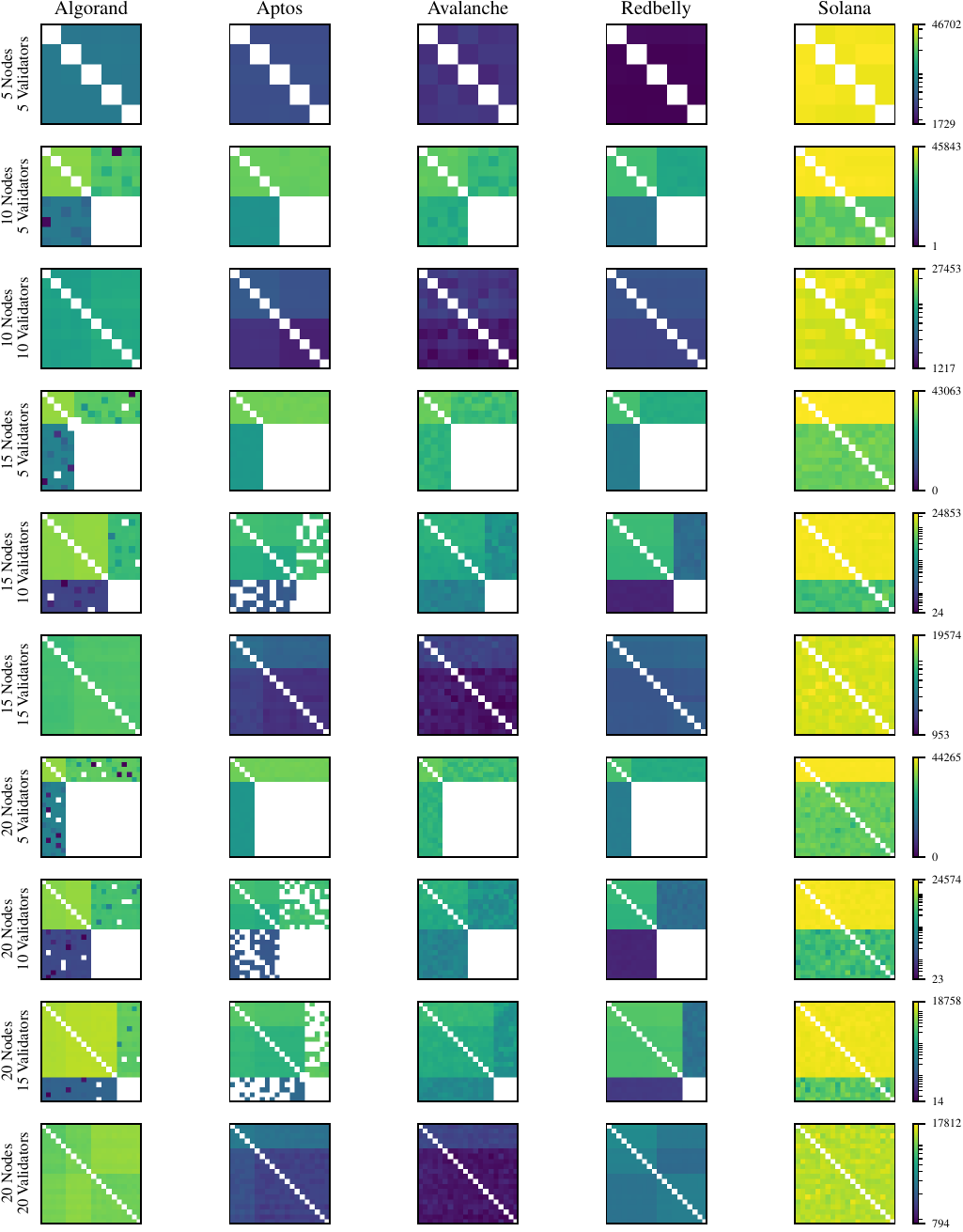}
   \caption{Heatmaps $M_{i,j}$ of the bandwidth used in \unit{\kibi\byte} between sending node $n_i$ and receiving node $n_j$ for each blockchain.\label{fig:total}}
\end{figure*}

\subsection{Large consumption variations depending on network scale}

\cref{fig:total} depicts a comprehensive grid of heatmaps for every blockchain across different combinations of number $N$ of nodes and number $V$ of validators. In each heatmap, the bandwidth consumed by node $n_i$ sending to node $n_j$ is represented by the color in cell $M_{i,j}$.

The raw data reveals significant disparities in the magnitude of data exchange across the different protocols. In a fully interconnected small network (5 nodes, all validators), the difference in per-link bandwidth is striking: while Redbelly and Avalanche maintain a lean footprint with approximately \qty{1700}{\kibi\byte} and \qty{2800}{\kibi\byte} transmitted per pair respectively, Algorand consumes considerably more, averaging around \qty{6500}{\kibi\byte} per link. Solana, however, operates at a completely different order of magnitude, with cells in the 5-node configuration consistently showing over \qty{42000}{\kibi\byte} of traffic---nearly 25 times the bandwidth usage of Redbelly for the same workload.

Furthermore, the data highlights a clear hierarchy of network load based on node roles. In mixed configurations (e.g., 20 nodes with 5 validators), the bandwidth intensity within the validator group (the top-left $5 \times 5$ sub-matrix) is drastically higher than the traffic involving non-validator nodes. For instance, in the 10-5 setup, for Aptos and Algorand, the traffic between two validators remains in the thousands of \unit{\kibi\byte} (e.g., \qty{\approx 3600}{\kibi\byte} for Aptos), whereas traffic originating from non-validator nodes often drops to the low hundreds (e.g., \qty{\approx 200}{\kibi\byte}), illustrating a highly centralized bandwidth burden on the consensus committee.

\subsection{Validators communicate more with themselves}

A distinct segmentation of the network topology is visible in most protocols. For Algorand, Aptos, Avalanche, and Redbelly, the top-left $V \times V$ sub-matrix is consistently dense and bright, confirming that validators communicate heavily among themselves to achieve consensus. However, the behavior regarding non-validator nodes varies significantly.

As observed, Solana is the unique outlier. It is the only blockchain where the non-validator nodes communicate directly with other non-validator nodes (the bottom-right quadrant of the heatmaps). For instance, in the 10-5 configuration, the heatmap is uniformly populated, indicating a full mesh topology where every node, regardless of its role, exchanges data with every other node.

In contrast, blockchains like Redbelly and Avalanche show a clear separation. While validators send data to non-validators (top-right quadrant) and receive data from them (bottom-left quadrant), non-validator nodes do not communicate with each other. This is evident in the 10-5 matrices for Avalanche and Redbelly, where the bottom-right $5 \times 5$ sub-matrix is largely empty.

\begin{figure*}[ht]
    \centering
    \includegraphics{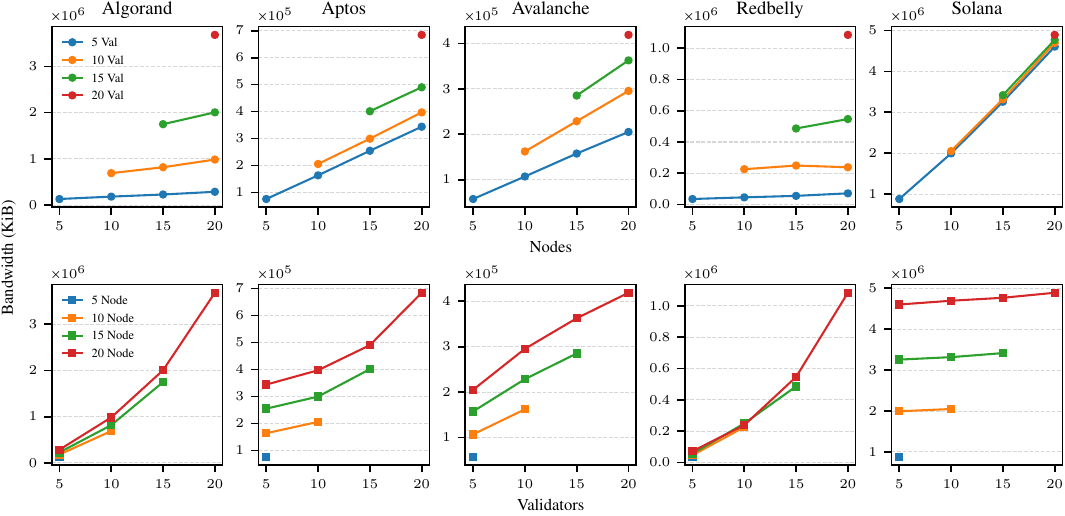}
   \caption{Total bandwidth consumption (\unit{\kibi\byte}) for blockchains under test. The top row plots bandwidth against the number of nodes (grouped by validator count), while the bottom row plots bandwidth against the number of validators (grouped by node count).\label{fig:trends}}
\end{figure*}

Finally, Algorand and Aptos exhibit the strictest separation in certain configurations. Non-validator nodes act almost exclusively as passive receivers or pull-based clients. In Algorand's 10-5 configuration, while validators send data to non-validators (rows 0-4 to columns 5-9), the traffic from non-validators back to validators is negligible, and traffic between non-validators is non-existent.

\subsection{Aptos pattern change with the number of validators}

\cref{fig:total} reveals a dynamic dissemination strategy in Aptos that depends on the ratio of validators to the total network size. In configurations where the number of validators is small relative to the network size, validators appear to broadcast to all non-validators. For example, in the 10-5 configuration and 20-5 configuration, the top-right quadrants are dense, showing that the 5 validators are sending data to all 5 (or 15) non-validator nodes.

However, as the number of validators increases, Aptos shifts strategy to reduce bandwidth overhead. In the 15-node, 10-validator configuration, the top-right quadrant becomes sparse. Specific validators only communicate with specific non-validator nodes rather than broadcasting to the entire set. This suggests a sharding or randomized gossip approach to dissemination when the validator set is large, likely to prevent bandwidth saturation on individual nodes.

\subsection{Blockchains vary with regards to bandwidth scalability}

 \cref{fig:trends} depicts the trends of bandwidth consumption depending on the scale, where each figure provides: line plots showing $S$ as a function of the total number $N$ of nodes, and line plots showing $S$ as a function of the number $V$ of validators. Here, $S$ represents the total sum of bandwidth measured in \unit{\kibi\byte} during the experiment duration.

We can clearly see from \cref{fig:trends} that $S_{Solana}$ bandwidth consumption increases primarily with the number of nodes, regardless of the number of validators. For instance, with 5 validators, increasing the total node count from 5 to 20 causes bandwidth to surge from \qty{879268}{\kibi\byte} to \qty{4604375}{\kibi\byte}---a five-fold increase. However, keeping the node count fixed at 20 and increasing validators from 5 to 20 results in a negligible increase from \qty{4604375}{\kibi\byte} to \qty{4891524}{\kibi\byte}. This confirms that Solana requires all nodes to communicate with each other, creating a high-bandwidth mesh topology dependent on the network size rather than the validator set size.

In contrast, blockchains $b \in \{Algorand, Redbelly\}$ the bandwidth $S_b$ depend almost entirely on the number of validators. Adding non-validator nodes adds very little overhead. For Algorand, with 5 validators, increasing nodes from 10 to 20 only increases usage from \qty{182790}{\kibi\byte} to \qty{287032}{\kibi\byte}. However, increasing validators has a massive impact: with 20 nodes, moving from 5 to 20 validators causes usage to skyrocket from \qty{287032}{\kibi\byte} to \qty{3681191}{\kibi\byte}---an increase of over 12 times. Redbelly exhibits an even more drastic ratio, jumping from \qty{72098}{\kibi\byte} (20 nodes, 5 validators) to \qty{1084185}{\kibi\byte} (20 nodes, 20 validators).

Finally, in two blockchains, Aptos and Avalanche, the bandwidth consumption grows with both the number of validators and the number of nodes. Note that the bandwidth increase appears superlinear with the number of validators in Aptos but sublinear in Avalanche. In Aptos, with 20 nodes, the bandwidth grows moderately between 5 and 15 validators (\qty{343968}{\kibi\byte} to \qty{490237}{\kibi\byte}) but jumps sharply when reaching 20 validators (\qty{684649}{\kibi\byte}). Conversely, Avalanche shows diminishing returns in bandwidth growth as validators are added to a 20-node network, rising from \qty{205154}{\kibi\byte} (5 validators) to \qty{418852}{\kibi\byte} (20 validators), indicating a more consistent propagation overhead.

Based on these trends, we can classify blockchains in three categories:
\begin{enumerate}
    \item {\bf Node-dependent (Solana)} Bandwidth consumption increases with the total network size ($N$), indicating a topology where all nodes propagate data heavily.
    \item {\bf Validator-dependent (Algorand and Redbelly)} Bandwidth consumption increases primarily with the size of the consensus committee ($V$). Non-validator nodes are passive consumers. %
    \item {\bf Hybrid (Avalanche and Aptos)} Bandwidth consumption increases with both $N$ and $V$, suggesting a topology where non-validator nodes participate in propagation. %
\end{enumerate}

\section{Consumption Variation with \unit{\TPS}}\label{sec:tps}

\begin{figure}[ht]
    \centering
        \centering
        \includegraphics{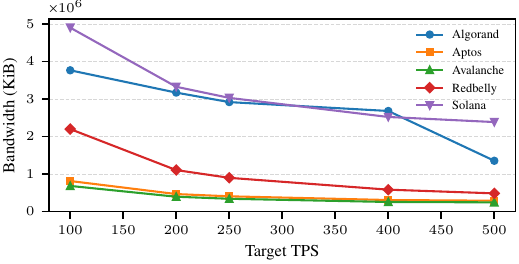}
        \caption{Total bandwidth consumption (\unit{\kibi\byte}) vs. Target \unit{\TPS} for blockchains under test under varying load rates. The workload was fixed at \num{20000} total transactions per run, with experiment duration scaling inversely to \unit{\TPS} (\qty{200}{\second} down to \qty{40}{\second}).\label{fig:bw-tps}}
\end{figure}

The bandwidth consumption could simply be a consequence of the rate at which our experiments send transactions. Below, we vary this sending rate to observe how it impacts bandwidth consumption. 
To this end, we conducted a set of experiments where the total workload size was kept constant at approximately \num{20000} transactions. We varied the target \unit{\TPS} from 100 to 500, inversely adjusting the experiment duration (from \qty{200}{\second} down to \qty{40}{\second}) to maintain the fixed transaction count ($\text{TPS} \times \text{Duration} \approx \text{Constant}$). \cref{fig:bw-tps} presents these results from the perspective of total absolute bandwidth. %

\cref{fig:bw-tps} plots the total absolute bandwidth consumed by each blockchain. A consistent trend is visible across all protocols: total bandwidth consumption decreases as \unit{\TPS} increases. Since the number of transactions is fixed, this decrease indicates that a significant portion of bandwidth usage is time-dependent rather than transaction-dependent. At lower \unit{\TPS} (longer duration), ``background'' traffic, such as heartbeats, empty block proposals, and consensus maintenance, accumulates, inflating the total footprint.

Solana consistently consumes the highest absolute bandwidth, ranging from approximately \qty{4.7}{\gibi\byte} (\qty{4905820}{\kibi\byte}) at \qty{100}{\TPS} down to \qty{2.3}{\gibi\byte} (\qty{2387814}{\kibi\byte}) at \qty{500}{\TPS}. Algorand follows as the second most bandwidth-intensive chain (\qty{3.6}{\gibi\byte} to \qty{1.3}{\gibi\byte}), while Avalanche and Aptos remain the most efficient, with Aptos consuming as little as \qty{283}{\mebi\byte} (\qty{289834}{\kibi\byte}) at \qty{500}{\TPS}. Notably, in terms of absolute bandwidth, Algorand consistently consumes more than Redbelly across all data points (e.g., at \qty{100}{\TPS}: Algorand \qty{\approx 3.6}{\gibi\byte} vs. Redbelly \qty{\approx 2.1}{\gibi\byte}).

\section{Related Work}\label{sec:rw}

\paragraph{Blockchain benchmarking}

Existing frameworks like Blockbench~\cite{DWC17} and Hyperledger Caliper~\cite{Kel24} standardize the evaluation of throughput, latency, and fault tolerance. Gromit~\cite{NDV23} further addresses ad-hoc testing limitations. However, these frameworks prioritize execution capacity (TPS) and treat the network layer as secondary. Our work builds upon Diablo~\cite{GGLNV23} and {\sc Stabl}~\cite{GGLV25} to rigorously isolate bandwidth consumption, a critical metric overlooked by execution-focused benchmarks.

\paragraph{Network traffic and block propagation}
Early studies analyzed propagation delays in Bitcoin~\cite{DW13}, leading to bandwidth optimization protocols like Graphene~\cite{OAL19}, Compact Blocks~\cite{Cor16}, and the BloXroute~\cite{KBK19} Layer-0 CDN. While these works propose techniques to minimize traffic, our paper provides a comparative measurement of modern Layer-1 protocols. We reveal that contemporary systems like Solana often prioritize data redundancy over the bandwidth efficiency emphasized in earlier Bitcoin and Ethereum research.

\paragraph{Communication complexity vs. empirical reality}
While theoretical literature favors linear communication complexity ($O(n)$) protocols like HotStuff~\cite{YMR19} over quadratic BFT implementations~\cite{castro_practical_1999}, theoretical bounds often fail to predict real-world behavior. By contrast, Voron and Gramoli~\cite{VG23} showed empirically that quadratic complexity can achieve significantly better performance when well balanced across a quadratic number of routes of a WAN. Di Perna et al.~\cite{DPBF25} showed that dense network topologies improve performance under load, a finding correlated with higher energy usage~\cite{DPSFB25}.
Our empirical results confirm this divergence between redundancy-heavy architectures like Solana and the lean traffic profiles of Redbelly.

\section{Conclusion}\label{sec:conclusion}

This paper compares the bandwidth consumption of blockchains for the first time. To this end, we measured empirically the traffic of five major layer-1 blockchains Algorand, Aptos, Avalanche, Redbelly, and Solana.

Our results show that transport protocols (polling vs. WebSockets) and block propagation strategies impact bandwidth more than transaction size.
Solana consumes up to 58\% more bandwidth than Algorand due to redundancy, while Aptos suffers from high idle overhead. Crucially, Solana scales with network size, whereas Algorand and Redbelly scale with validator count.

\section*{Acknowledment}

This
research is supported under Australian Research Council Discovery Project funding scheme (project number 250101739) entitled ``Fair Ordering of Decentralised Access to Resources''.

\bibliographystyle{plainurl}
\bibliography{references}

\end{document}